\documentclass[10pt,journal,twoside]{IEEEtran}

\ifCLASSOPTIONcompsoc
  \usepackage[nocompress]{cite}
\else
  \usepackage{cite}
\fi
\ifCLASSINFOpdf
   \usepackage[pdftex]{graphicx}
\else
   \usepackage[dvips]{graphicx}
\fi
\usepackage{amsmath}
\usepackage{acronym}
\usepackage{algorithmic}
\usepackage{array}
\ifCLASSOPTIONcompsoc
 \usepackage[caption=false,font=footnotesize,labelfont=sf,textfont=sf]{subfig}
\else
 \usepackage[caption=false,font=footnotesize]{subfig}
\fi
\usepackage{fixltx2e}
\usepackage{url}

\newcommand\MYhyperrefoptions{bookmarks=true,bookmarksnumbered=true,
pdfpagemode={UseOutlines},plainpages=false,pdfpagelabels=true,
colorlinks=true,linkcolor={black},citecolor={black},urlcolor={black},
pdftitle={Nayem_DraftTASLP},
pdfsubject={Typesetting},
pdfauthor={Khandokar Md. Nayem},
pdfkeywords={TASLP, journal, LaTeX, paper,
             template}}
\ifCLASSINFOpdf
\usepackage[\MYhyperrefoptions,pdftex]{hyperref}
\else
\usepackage[\MYhyperrefoptions,breaklinks=true,dvips]{hyperref}
\usepackage{breakurl}
\fi

\usepackage{booktabs}
\usepackage{multirow}
\usepackage{adjustbox, bm}
\usepackage{xcolor}
\usepackage{pgfplots}
\pgfplotsset{compat = 1.9}
\usepgfplotslibrary{colorbrewer}
\usetikzlibrary{pgfplots.statistics, pgfplots.colorbrewer} 
\usepackage{tikz}
\usepackage{amsfonts}
\usepackage{pgfplotstable}
\usepackage{filecontents}

\DeclareMathOperator*{\argmax}{arg\,max\,}

\begin{document}
%
\title{Attention-based Speech Enhancement Using Human Quality Perception Modelling}
%
%
\author{Khandokar~Md.~Nayem,~\IEEEmembership{Student Member,~IEEE,}
        and~Donald~S.~Williamson,~\IEEEmembership{Senior Member,~IEEE}

\thanks{This work was supported in part by NSF under Grant IIS-1942718 and in part by Lilly Endowment, Inc., through its support for the Indiana University Pervasive Technology Institute}%
\thanks{Khandokar Md. Nayem is with the Department of Computer Science, Indiana University, Bloomington, IN 47408 USA (e-mail: knayem@iu.edu).}%

\thanks{Donald S. Williamson was with the Department of Computer Science at Indiana University, but he is now with the Department of Computer Science and Engineering, Ohio State University, Columbus, OH, USA 43210 USA (e-mail: 
williamson.413@osu.edu).}%

}


\IEEEtitleabstractindextext{%
\begin{abstract}
Perceptually-inspired objective functions such as the perceptual evaluation of speech quality (PESQ), signal-to-distortion ratio (SDR), and short-time objective intelligibility (STOI), have recently been used to optimize performance of deep-learning-based speech enhancement algorithms. These objective functions, however, do not always strongly correlate with a listener's assessment of perceptual quality, so optimizing with these measures often results in poorer performance in real-world scenarios. In this work, we propose an attention-based enhancement approach that uses learned speech embedding vectors from a mean-opinion score (MOS) prediction model and a speech enhancement module to jointly enhance noisy speech. The MOS prediction model estimates the perceptual MOS of speech quality, as assessed by human listeners, directly from the audio signal. The enhancement module also employs a quantized language model that enforces spectral constraints for better speech realism and performance. We train the model using real-world noisy speech data that has been captured in everyday environments and test it using unseen corpora. The results show that our proposed approach significantly outperforms other approaches that are optimized with objective measures, where the predicted quality scores strongly correlate with human judgments. 
\end{abstract}

}

\maketitle

\IEEEdisplaynontitleabstractindextext
\IEEEpeerreviewmaketitle


\ifCLASSOPTIONcompsoc
\IEEEraisesectionheading{\section{Introduction}\label{sec:introduction}}
\else
\section{Introduction}
\label{sec:introduction}
\fi

\IEEEPARstart{M}{onaural} speech enhancement aims to remove unwanted noise from an audio signal that contains speech using only a single microphone channel. Enhancing the quality of noisy speech is crucial for applications such as speech recognition, speaker verification, hearing aids, and hands-free communication. Speech enhancement approaches are generally divided into two categories: mask-based or signal-based approximation. A time-frequency (T-F) mask is estimated in mask-based approaches, where the mask filters unwanted noise from noisy speech mixtures. Early mask-based approaches estimate the ideal binary mask (IBM)~\cite{li2008factors} or the ideal ratio mask (IRM)~\cite{narayanan2013ideal}, while recent approaches estimate the phase-sensitive mask (PSM)~\cite{erdogan2015phase} or complex ideal ratio mask (cIRM)~\cite{williamson2015complex, lee2019joint} to enhance both the magnitude and phase. The ideal quantized mask (IQM) has recently been proposed~\cite{healy2018ideal}, where each T-F unit of the IRM is assigned to a quantization level according to its signal-to-noise ratio. It has been shown to be a reasonable representation of the IRM as assessed by human listeners, however, estimation of the IQM and its subsequent noise removal has not be thoroughly evaluated. 

Signal approximation can be done in either the time~\cite{luo2018tasnet, pandey2019new} or the T-F domains~\cite{odelowo2018study}, where the approach directly estimates the time or T-F domain signal from a noisy speech representation. Traditionally, T-F masks produce better objective quality and intelligibility compared to direct signal approximation, mainly because masks are normalized and bounded with limited speaker variations, which makes them easier to learn. Also, masks directly modulate the mixture signal in the T-F domain. In recent years, the signal approximation models outperform mask estimation approaches in speech intelligibility~\cite{odelowo2018study, lu2022conditional} when applied with appropriate normalization.

Regardless of the approach, recent developments in deep learning have resulted in state-of-the-art performance. A wide range of deep learning architectures have been proposed, including, deep neural networks (DNNs)~\cite{xu2013experimental, wang2014training}, autoencoders~\cite{xia2013speech,lu2014ensemble,lee2016two}, long short-term memory (LSTM) networks~\cite{weninger2014discriminatively,erdogan2015phase}, convolutional neural networks (CNNs)~\cite{zhao2018convolutional, tan2018convolutional, choi2018phase, pandey2019new, kolbaek2020loss}, and generative adversarial networks (GANs)~\cite{pascual2017segan, donahue2018exploring, fu2019metricGAN}. Deep recurrent networks have proven to be effective, especially compared to fully-connected DNNs, as they capture temporal correlations. 
CNNs are good at feature extraction, and they have been combined with recurrent networks to capture the short and long-term temporal and spectral correlations. Recently, attention-based deep architectures have been proposed with the motivation that a training target only greatly influences a few regions of the input, where the focal regions change over time. \cite{giri2019attention, tolooshams2020channel} use attention mechanism with an U-Net~\cite{ronneberger2015u} architecture for both time and spectral domain speech enhancement. \cite{hao2019attention, koizumi2020speech} successfully use self-attention to estimate a speech spectrum and T-F mask, respectively. This approach is more intuitive for speech enhancement, because humans are able to focus on the target speech with high attention while paying less attention to the noise. 
 
Deep-learning-based speech enhancement approaches traditionally use the mean square error (MSE) between the short-time spectral-amplitudes (STSA) of the estimated and clean speech signals to optimize performance. This is done due to the computational efficiency of the MSE loss function. However, the MSE tends to produce overly-smoothed speech and it is not always a strong indicator of performance~\cite{wang2009mean, shu2020human}. Thus, many studies have begun to optimize algorithms using perceptually-inspired objective measures. 
 
Multiple studies have used short-time objective intelligibility (STOI)~\cite{taal2011algorithm} to optimize enhancement algorithms and to improve speech intelligibility~\cite{zhang2018training, fu2018end, koizumi2018dnn}. This is done to minimize the inconsistency between the model optimization criterion and the evaluation criterion for the enhanced speech. The reported results in \cite{fu2018end} show that jointly optimizing with STOI and MSE improves speech intelligibility according to both objective and subjective measures. In addition, word accuracy according to automatic speech recognition (ASR) is improved. Perceptual evaluation of speech quality (PESQ)~\cite{rix2001perceptual} scores, however, have not increased when optimizing with STOI, as reported in~\cite{fu2018end}. The signal-to-distortion ratio (SDR)~\cite{fevotte2005bss_eval} has also been used as an objective cost function~\cite{kawanaka2020stable}. The proposed network is pre-trained with a SDR loss to achieve network stability and later optimized with a PESQ loss in a black-box manner. Their results show that optimizing with SDR leads to overall objective quality improvements. Unlike SDR and STOI, PESQ cannot directly be used as an objective function since it is non-differential. Reinforcement learning (RL) techniques such as deep Q-network and policy gradient have thus been employed to solve the non-differentiable problem \cite{koizumi2018dnn, koizumi2017dnn}. In these works, PESQ and the perceptual evaluation methods for audio source separation (PEASS)~\cite{emiya2011subjective, vincent2012improved} serve as rewards that are used to optimize model parameters. Meanwhile, a new PESQ-inspired objective function that considers symmetrical and asymmetrical disturbances of speech signals has been developed in~\cite{martin2018deep}. Quality-Net~\cite{fu2018quality}, which is a DNN approach that estimates PESQ scores given a noisy utterance, has also been used as a maximization criteria~\cite{fu2019learning} and as a model selection parameter~\cite{zezario2019specialized} to enhance speech. 

It is worth noting that optimizing with perceptually-inspired objective measures has been disputed in \cite{kolbaek2018monaural, kolbaek2018relationship}, where these latter results show that a MSE objective function is sufficient. This may occur because objective measures of success do not always strongly correlate with subjective measures~\cite{emiya2011subjective, rix2006objective,cano2016evaluation, santos2014improved}. Hence, it is inconclusive as to whether perceptually-inspired objective measures are generally useful at optimizing speech enhancement performance, so alternative strategies for incorporating perceptual feedback may be needed.

Subjective evaluations from human listeners remains the gold-standard approach since it results in ratings from potential end-users. These evaluations often ask listeners to either give relative preference scores~\cite{quackenbush1988objective} or assign a numerical rating~\cite{malfait2006p}. Multiple ratings are provided for each signal, where they are averaged to generate a mean-opinion score (MOS). Recently, deep-learning approaches have effectively estimated human-assessed MOS~\cite{avila2019non, patton2016automos, lo2019mosnet, dong2020attention}. These approaches are promising since they can provide strongly correlated quality scores for new signals. According to \cite{braun2022effect}, a non-intrusive loss function can lead to improved noise suppression. Conversely, \cite{zezario2022deep} proposes using embedding vectors from a multi-objective speech assessment model for speech enhancement, but they only use intrusive metrics such as PESQ, STOI, and a speech distortion index (SDI) to train the speech assessment model. As a result, it remains unclear whether a speech assessment model that predicts MOS can incorporate human perceptual information into a speech enhancement model.

Joint learning has been successfully applied in speech enhancement to optimize between estimating speech and other training targets, such as phoneme classification~\cite{schulze2020joint}, speaker identification~\cite{ji2020speaker}, and speech recognition~\cite{donahue2018exploring}. Our preliminary work has recently combined a speech quality estimation task with speech enhancement~\cite{nayem2021incorporating} and it shows promising results. In this work, we propose an attention-based speech enhancement model that uses the embedding vector from a MOS prediction model to produce speech with improved perceptual quality. The MOS estimator generates encoded embedding vectors that contain perceptually useful information that is important for human-based assessment. Our speech enhancement attention model is conditioned on that embedding vector and enhances the noisy speech using a separate encoder-decoder framework, which should help produce better quality speech according to human evaluation.  In the enhancement stage, we incorporate a quantized spectral language model that captures the transitions probabilities across the T-F spectrum. The LM helps ensure that the resulting speech spectra exhibit realistic spectral- and temporal-fine structure that occurs within real speech signals, since it identifies the most likely spectrum in each time frame. This is accomplished by first quantizing the speech magnitude spectra into distinct classes. Our proposed signal approximation approach jointly updates both the MOS-prediction and speech-enhancement models during training, using speech enhancement and MOS prediction loss terms.



\begin{figure*}[tbh!]
    \centering
    \includegraphics[width = 0.8\linewidth]{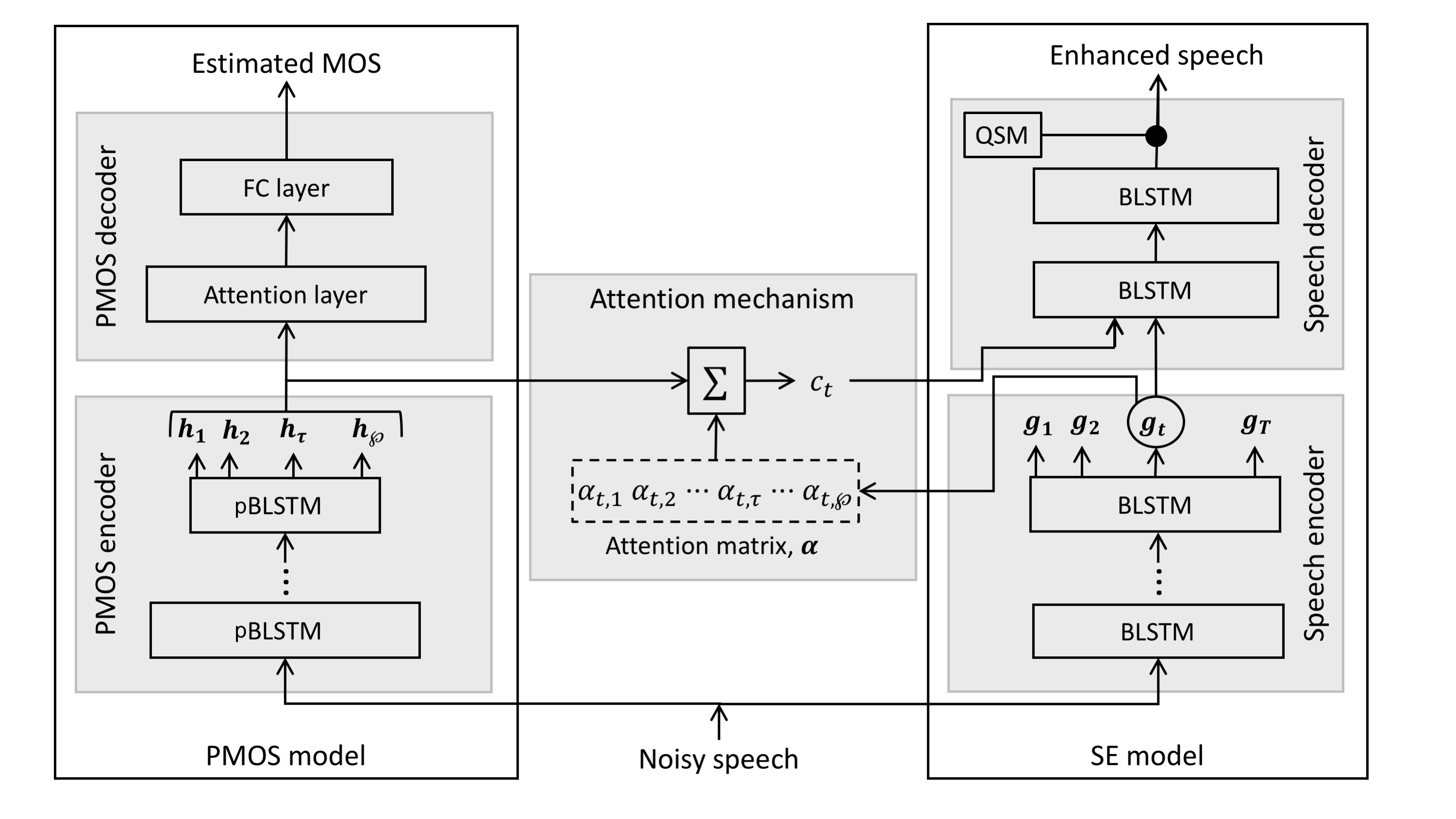}
    \caption{A depiction of our speech-enhancement model that consists of a MOS-prediction model denoted as PMOS (left side), and a speech-enhancement (SE) model (right side). An attention mechanism connects the two models.}
    \label{fig:model}
\end{figure*}

The rest of the paper is organized as follows. In section~\ref{sec:methods}, we introduce the quality assessment model, the proposed enhancement model, and the quantized spectral language model. We describe our dataset and experimental setup in section~\ref{sec:experiments}. In section~\ref{sec:results}, we evaluate our proposed approach and compare it with other state-of-the-art models. We discuss the implication and significance of our work in section~\ref{sec:discuss}. Finally, we conclude our work in section~\ref{sec:conclusion}.


\section{Proposed Approach}
\label{sec:methods}

A depiction of our approach is shown in Figure \ref{fig:model}. The model consists of a MOS prediction model (shown left) and a speech enhancement model (shown right). Our MOS prediction model is tailored to provide estimates for subjective-MOS (as rated by humans), and going forward, we will use MOS to refer to subjective-MOS unless explicitly stated otherwise, for ease of understanding. We next will provide notation and then describe each of these sub-modules.

\subsection{Notation}

We define a clean speech signal as $s_t$ and background noise as $n_t$ at time $t$. The mixture of clean speech and noise is denoted as $m_t=s_t+n_t$. We aim to extract the speech from the mixture by removing the unwanted noise. The short-time Fourier transform (STFT) converts the time-domain mixture into a T-F representation, $M_{t,f}$, that is defined at time $t$ and frequency $f$. The complex-valued STFT matrix, $\bm{M}$, can be written as $\bm{M}=|\bm{M}|e^{i\bm{\theta}^M}$ with magnitude $|\bm{M}|\in \bm{\Re}^{T\times F}_+$ and phase $\bm{\theta}^M \in \bm{\Re}^{T\times F}$, where $T$ is the length of speech in time and $F$ is the total number of frequency channels.

Enhancing the magnitude response of noisy speech results in an estimate of the clean speech magnitude response, $|\hat{\bm{S}}|$, using an enhancement function $\mathcal{F}_\delta$ such that $|\hat{\bm{S}}| =\mathcal{F}_\delta(|\bm{M}|)$. The enhancement function is modeled with a deep neural network which is trained to maximize the conditional log-likelihood of the training dataset, 
\begin{align*}
    &\max \frac{1}{N} \sum^N \log P\Big( |{\bm{S}}| \, \Big| \, |\bm{M}|\Big) \\
    \Rightarrow &\max_\delta \frac{1}{N} \sum^N \log P\Big( \mathcal{F}_\delta(|\bm{M}|) \, \Big| \, |\bm{M}|\Big)
\end{align*}
where $\delta$ denotes the set of tunable parameters and $N$ is the number of training examples. The estimated magnitude response $|\hat{\bm{S}}|$ is then combined with the noisy phase, $\bm{\theta}^M$, where the inverse STFT produces an enhanced speech signal in the time domain, $\hat{s}_t$. 

\subsection{Speech quality assessment model}
\label{subsec:mos_model}

A MOS prediction model proposed by \cite{dong2020pyramid} is adapted to estimate the MOS from noisy speech. This model has been developed with real-world captured data and it has been shown to outperform comparison approaches~\cite{fu2018quality, avila2019non, mittag2019non}, according to multiple metrics. The MOS prediction model consists of an attention-based encoder-decoder structure that uses stacked pyramid bi-directional long-short term memory (pBLSTM)~\cite{chan2016listen} networks in the encoder. We denote this model as Pyramid-MOS (PMOS). A pBLSTM architecture gives the advantages of processing sequences at multiple time resolutions, which effectively captures short- and long-term dependencies. Speech has spectral and temporal dependencies over short and long durations, and a multi-resolution framework is effective in learning these complex relations.

\begin{figure}[t!]
    \centering
    \includegraphics[width = 0.95\linewidth]{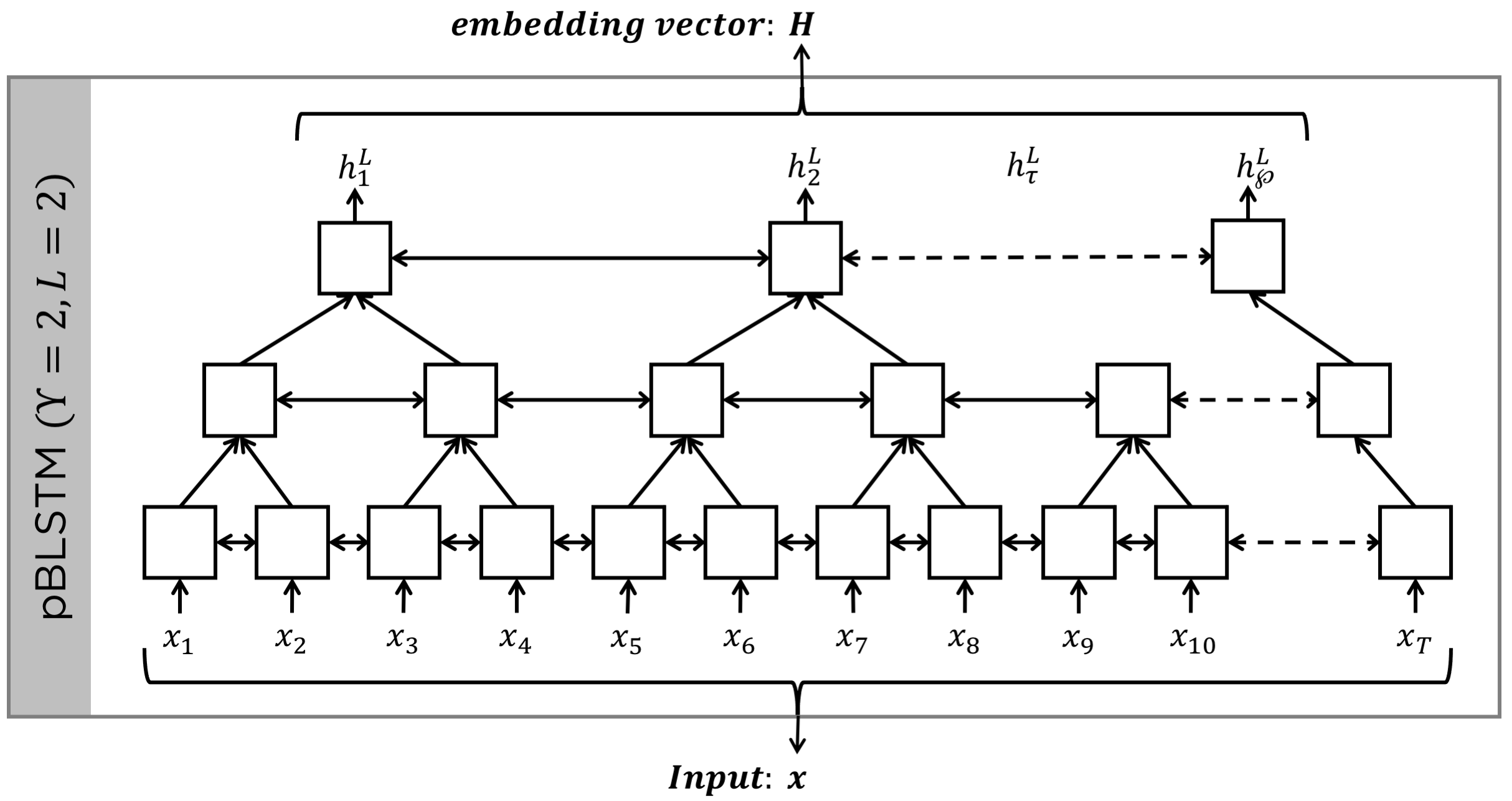}
    \caption{Illustration of pBLSTM structure with reduction factor $\Upsilon=2$ and number of layer $L=2$.}
    \label{fig:pBLSTM}
\end{figure}

A single T-F frame of the noisy-speech mixture, $|\bm{M}_t|$, is the input to the PMOS encoder. In a pyramid structure, the lower layer outputs from $\Upsilon$ consecutive time frames are concatenated and used as inputs to the next pBLSTM layer, along with the recurrent hidden states from the previous time step. The output of a pBLSTM node is an embedding vector, $h^l_t$, that is as defined below:
\begin{align}
    h^l_t &= pBLSTM\Big( h^l_{t-1}, \big[ h^{l-1}_{\Upsilon\times t -\Upsilon+1}, h^{l-1}_{\Upsilon\times t}\big] \Big)
\end{align}
where $\Upsilon$ is the reduction factor (e.g., number of concatenated frames) between successive pBLSTM layers and $l$ is the layer number. A pBLSTM reduces the time resolution from the input speech to the final latent representation $\bm{H}$. Figure~\ref{fig:pBLSTM} shows the internal structure of pBLSTM module.
This compressed vector accumulates the useful features for measuring speech perceptual quality that resides in a range of time-frames and ignores the least important features.
The encoder outputs a concatenated version of the hidden states of the last pBLSTM layer as vector $\bm{H}=\{\bm{h}_1, \dotsb, \bm{h}_\tau, \dotsb, \bm{h}_\wp\}$, where $\wp$ is the total number of final embedding vectors with time index $\tau$.

The output of the PMOS encoder becomes the input to the PMOS decoder unit. This decoder is implemented as an attention layer followed by a fully-connected (FC) layer and it outputs an estimated MOS of the input speech utterance. Attention models learn key attributes of a latent sequence, since adjacent time frames can provide important information, which is particularly crucial for our task.  
The attention mechanism~\cite{luong2015effective} uses the pyramid encoder output at the $i$-th and $k$-th time steps to compute the attention weights, $\alpha^{PMOS}_{i,k}$. Attention weights are used to compute a context vector, $c^{PMOS}_i$, using the following equations:
\begin{align}
    \alpha^{PMOS}_{i,k} &= \frac{\exp{(\bm{h}_i^\top \bm{Q} \bm{h}_k)}}{\sum^{\wp}_{\phi=1} \exp{(\bm{h}_i^\top \bm{Q} \bm{h}_\phi)}}\\
    c^{PMOS}_i &= \sum^\wp_{k=1} \alpha^{PMOS}_{i,k} \cdot \bm{h}_k
\end{align}
$\bm{Q}^{\wp\times\wp}$ is the trainable PMOS attention weight matrix. We learn $\bm{Q}$ using a feed-forward neural network that attempts to capture the alignment between the embeddings $\bm{h}_i$ and $\bm{h}_k$. 

The context vector is provided to a fully-connected layer to estimate the MOS. Note that the pyramid structure of the encoder results in a shorter sequence of latent representations than the original input sequence, and it leads to fewer encoding states for attention calculation at the decoding stage. Therefore, strictly  $\wp<T$, and in our case $\wp = \lceil T/\Upsilon^L \rceil$, where $L$ is the number of pBLSTM layers.
We train the PMOS model separately with the parameters defined in~\cite{nayem2019incorporating}. After training, this model is held frozen during inference.

\subsection{Proposed speech enhancement model}
\label{subsec:se_model}
Our proposed speech-enhancement (SE) model follows an encoder-decoder structure, and it is shown in Figure \ref{fig:model} (right). The SE encoder takes a single T-F frame of a noisy-speech mixture, $|\bm{M}_t|$, as input and multiple BLSTM layers, are stacked together to create a hidden representation of the frame, $\bm{g}_t$. In our SE encoder, we utilize BLSTM layers instead of pBLSTM layers since we aim to estimate an embedding frame for each time frame and pBLSTM layers reduce the number of output frames. 
An attention mechanism is applied using the mixture encoding from the SE model, $\bm{G}=\{\bm{g}_1, \bm{g}_2, \dotsb, \bm{g}_T\}$, and the PMOS encoding, $\bm{H}$, from the MOS prediction model. This allows the SE model to exploit the MOS estimator's encoding and utilize the important perceptual feature embedding that correlates with human assessment. Considering that the pBLSTM structure of the PMOS encoder condenses the final encoding vector $\bm{H}$ along time, PMOS yields a smaller time resolution than the encoding from the SE encoder, so we compute a score for each embedding vector $\bm{h}_{\tau}$ using an alignment  weight matrix, $\bm{W}^{T\times\wp}$. Then the attention weights for the SE model, $\alpha_{t,\tau}$, are obtained using a softmax operation over the scores of all $\bm{h}_\tau$. Now, the PMOS encoding is summarized in a context vector $\bm{c}_t$ for each mixture frame $\bm{g}_t$. Prior to computing $\bm{c}_t$, $\bm{h}_\tau$ passes through a linear layer $\ell$, so that we learn a different representation for the SE task. The computations are below:
\begin{align}
    \alpha_{t,\tau} &= \frac{\exp{(\bm{g}_t^\top \bm{W} \bm{h}_\tau})}{\sum^{\wp}_{\phi=1} \exp{(\bm{g}_t^\top \bm{W} \bm{h}_\phi)}} \\
    \bm{c}_t &= \sum_{\tau=1}^\wp \alpha_{t,\tau} \cdot \ell (\bm{h}_\tau)
\end{align}
\noindent
Then, the context vector and SE-model embedding vector are concatenated (e.g., $[\bm{c}_t, \bm{g}_t]$) and passed to the decoder module. The SE-decoder module follows the network structure from \cite{schulze2020joint}. It consists of a linear layer with a $tanh(\cdot)$ activation function, two BLSTM layers, and a linear layer with ReLU activation. It outputs the estimated enhanced speech $|\hat{\bm{S}}|$. This estimated speech magnitude with noisy phase produce the estimated clean speech, i.e. $\hat{\bm{S}} = |\hat{\bm{S}}|e^{i\bm{\theta}^M}$. Since we are estimating two targets MOS and enhanced speech simultaneously, the unified model will learn different representations for these tasks. Thus both PMOS and SE models will learn their corresponding targets with perceptual feature sharing. We freeze the PMOS model while training this SE model.

\subsection{Joint-learning of PMOS and SE model}
\label{subsec:joint_model}
We also develop an approach that allows the PMOS and SE models to be jointly trained. Our joint-learning objective function uses a weighted average of a {time-domain} signal-approximation loss $\mathcal{L}_{sa}$ (from the SE model), the MSE of the magnitude spectrum $\mathcal{L}_{mse}$ (from the SE model) and the MSE of the MOS estimation $\mathcal{L}_{mos}$ (from the PMOS model). We compute the signal-approximation loss from the time-domain signal difference between the reference speech $s$ and enhanced speech $\hat{s}$. The overall loss function of our network is defined as below, using hyper-parameters $\lambda_1$ and $\lambda_2$ that control the impact of individual loss terms:
\begin{align}
    \mathcal{L} &= \lambda_1\left[\lambda_2\mathcal{L}_{mse} + (1-\lambda_2)\mathcal{L}_{sa}\right] + (1-\lambda_1)\mathcal{L}_{mos}
    \label{eq:loss}
\end{align}
\noindent
The model training order is as such. First, we train the PMOS model using $\mathcal{L}_{mos}$ (e.g. $\lambda_1 = 0$). Then we train the SE model using $\lambda_1 = 1$, while running the PMOS model in inference mode (e.g. it is held fixed). This is done to ensure that the trained PMOS model effectively encodes the key features in the embedding vector that are important to perceptual speech quality. Finally, we train both the models jointly (e.g. $0 < \lambda_1 < 1$) using $\mathcal{L}$ to further reduce any correctional differences between the true and estimated MOS in the PMOS model, and to increase the perceptual quality of the enhanced speech.
\begin{figure}[t!]
    \centering
    \includegraphics[width = 0.8\linewidth]{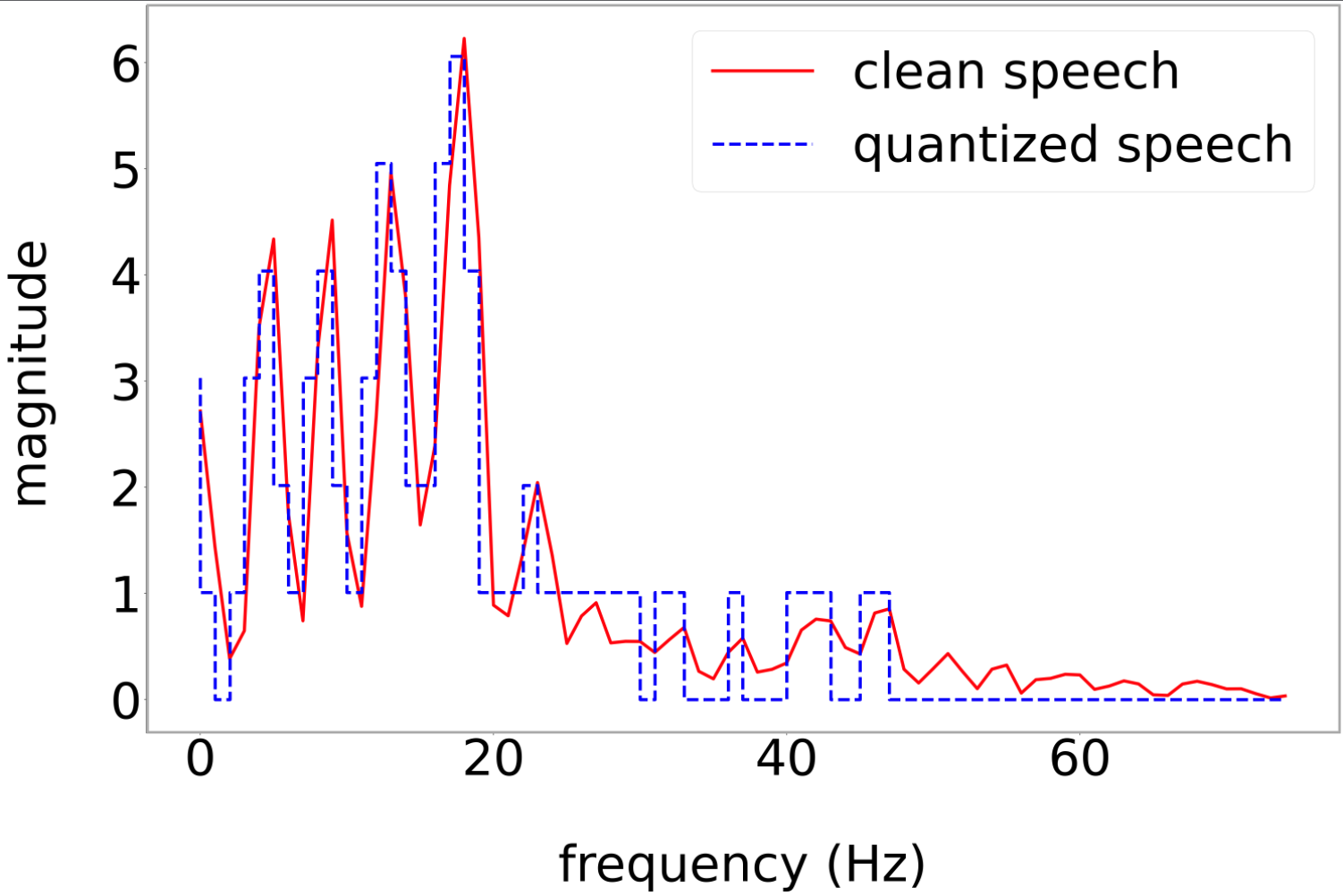}
    \caption{Quantization of a clean magnitude spectrum.}
    \label{fig:quant}
\end{figure}
\subsection{Quantized Spectral Model}
\label{subsec:QSM}
From written and spoken language, we can determine the sequences of words that are most likely to occur. This knowledge is captured by a language model (LM) of an automatic speech recognition system which we can expressed as,
\begin{align}
    \hat{words}=\argmax_{words\in Language} \overbrace{P(input|words)}^{acoustic\;model} \overbrace{P(words)}^{language\;model}
\end{align}
The LM is useful in eliminating rare and grammatically incorrect word sequences, and it enhances the performance of ASR systems. In the case of speech enhancement, models learn spectral information within frames over time, but they often neglect the temporal correlations. Our approach, as proposed in \cite{nayem2021towards}, suggests incorporating a ``LM" to fuse temporal correlations and overcome this limitation. Therefore, we construct a bi-gram Quantized Spectral Model (QSM), which functions in a similar way to a language model (LM), in order to produce more realistic spectra. The QSM estimates the probability of spectral magnitudes along time for each frequency channel conditioned on its previous T-F spectral magnitude. 
On a reference speech corpora, we apply a normalization scaling function, $\mathcal{N}_{[o,r]}(\cdot)$, that normalizes the magnitude spectrogram and re-scales the range to $[0,r]$. Then a quantization function, $\mathcal{Q}_\chi(\cdot)$, converts the range constrained magnitude spectrogram into $\mathcal{D}$ number of bins that are $\chi$ steps apart. This produces quantized speech, i.e. $|S|^q = \mathcal{Q}_\chi\big(\mathcal{N}_{[0,r]}(|S|)\big)$. Fig.~\ref{fig:quant} shows an example of the original clean and quantized clean magnitude spectra, where $\chi=2$ for display purposes. Our proposed QSM has $\mathcal{D}$ spectral levels. We construct the QSM using quantized speech magnitudes from the clean speech corpora. The QSM is less likely to suffer from the out of vocabulary problem when the model parameters, $\chi$ and $r$, are adequately defined.


We compute per-frequency-channel QSMs along the time axis where each entry, $d$, refers to a quantization attenuation level. We then compute the transition probability between two time consecutive T-F units, $fQSM_f = P(d_{t+1,f}|d_{t,f})$. The probabilities are calculated by counting the level transitions, and then normalizing by the appropriate scalar. These probabilities are stored in the per-frequency-channel QSM resulting in a $F\times \mathcal{D}\times \mathcal{D}$ probability matrix. 
We re-evaluate the transition probabilities using Good-Turing smoothing~\cite{jurafskyMartin2009} to overcome the zero-probability problem in N-grams. Shallow fusion~\cite{gulcehre2015using} is a simple method to incorporate an external LM into an encoder-decoder model, and it produces better results compared to others. Hence, we use shallow fushion to combine our QSM and SE model based on log-linear interpolations at inference time. This is shown in the below equations:

\begin{align}
    P^{QSM}_f(|\hat{\bm{S}}_{:,f}|) &= \prod^T_{i=1} P(d_{i,f}|d_{i-1,f}) \\
    |\hat{\bm{S}}_{:,f}|^* = \argmax_{|\hat{\bm{S}}_{:,f}|} &\log P\big(|\hat{\bm{S}}_{:,f}| \big| |\bm{M}|\big) + \mu \log P^{QSM}_f\big(|\hat{\bm{S}}_{:,f}| \big)
    \label{eq:S_hat}
\end{align}
\noindent
Here $P^{QSM}_f$ denotes the transitional probability of QSM at frequency $f$, $P\big(|\hat{\bm{S}}_{:,f}| \big| |\bm{M}|\big)$ represents the estimated magnitude output of the LSTM layers of the SE decoder, and $\mu$ is a hyper-parameter that is tuned to maximize the performance on a development set. Note that we train our QSM in advance on a clean speech corpus and use it in inference mode during enhancement. The tunable parameter $\mu$ of (\ref{eq:S_hat}) is set to zero when we do not have a trained QSM.


\section{Experiments}
\label{sec:experiments}

\subsection{Dataset}
\label{subsec:dataset}

We use the COnversational Speech In Noisy Environments (COSINE)~\cite{stupakov2009cosine} and the Voices Obscured in Complex Environmental Settings (VOiCES)~\cite{richey2018voices} corpora. 
COSINE captures multi-party conversations on open-ended topics for spontaneous and natural dialogue. These conversations are recorded in real world environments in a variety of background settings. The audio recordings are captured using 7-channel wearable microphones that consist of a close-talking mic (e.g., near the mouth, clean reference), far-field mic (on the shoulder), throat mic, and an array of four mics (spaced 3 cm apart) positioned in front of the speaker's chest. In total, 133 English speakers record 150 hours of audio with the approximated signal-to-noise ratios (SNR) ranging from -10.1 to 11.4 dB.

VOiCES contains audio recorded using 12 microphones placed throughout real rooms of different size and acoustic properties. Various background noises like TV, music, or babble are simultaneously played with foreground clean speech, so the recordings contain noise and reverberation. A foreground loudspeaker moves through the rooms during recording to imitate human conversation. This foreground speech is used as the reference clean signal, and the audio captured from the microphones is used as the reverberant-noisy speech. The approximate speech-to-reverberation ratios (SRRs) of the VOiCES signals range from -4.9 to 4.3 dB. 


The MOS data was collected from a listening study in \cite{dong2020pyramid}. Listeners assessed the speech quality of audio signals using a 100-point scale. In total, 45 hours of speech and 180k subjective human ratings are summarized into the MOS quality ratings for 18000 COSINE signals and 18000 VOiCES signals. The collected responses are processed further to mitigate rating biases~\cite{zielinski2008some}, remove responses that were unanswered or randomly scored~\cite{gadiraju2015understanding}, and to deal with outliers~\cite{ester1996density, liu2008isolation}. Z-score pruning~\cite{han2011data} followed by min-max normalization is performed, resulting in a MOS rating scale of 0 to 10.  The scaled ratings for each audio signal are finally averaged.

We additionally evaluate using the 4th CHiME Speech Separation and Recognition Challenge (CHiME-4)~\cite{vincent2017analysis} and the 5th CHiME Speech Separation and Recognition Challenge (CHiME-5)~\cite{barker2018fifth} corpora. We use these to investigate the generalization capacity of our proposed approach.

\subsection{System Setup}
\label{subsec:setup}

All signals are downsampled to $16$ kHz. Noisy or reverberant stimuli of each dataset are divided into training (70\%), validation (10\%), and testing (20\%) sets, and trained separately.

For MOS prediction, the input signals are segmented into 40 ms length frames with 25\% overlap. A 512-point FFT and a Hanning window are used to compute the spectrogram. Mean and variance normalization are applied to the input feature vector. The PMOS encoder consists of $256$ nodes followed by 3 pBLSTM layers ($L = 3$) with 128, 64 and 32 nodes in each direction, respectively. Like \cite{dong2020pyramid, chan2016listen}, the reduction factor $\Upsilon = 2$ is adopted here. As a result, the final latent representation $\bm{h}_\tau$ is reduced in the time resolution by a factor of $\Upsilon^3 = 8$. The outputs of two successive BLSTM nodes are fed as input to a BLSTM node in the upper layer. 
In the PMOS decoder, the context vector is passed to a fully connected (FC) layer with 32 units. The model is optimized using Adam optimization~\cite{kingma2015adam} with convergence determined by a validation set. Early stopping with initial learning rate of $0.001$ is applied in the training phase.

The proposed SE model uses a 640-point DFT with a Hann window of 40ms and a 20ms frame shift to generate the spectrogram for the encoder input. The SE encoder consists of 2 BLSTM recurrent layers. The SE decoder has a linear layer with $tanh$ activation, followed by 2-layers of BLSTM and a linear layer with ReLU activation~\cite{schulze2020joint, schulze2019weakly}. Each BLSTM layer contains 200 nodes and each linear layer has 321 nodes. The same optimization technique with early stopping by validation set is applied. 

For our proposed QSM language model, we choose a quantization step of $\chi=0.0625$, which was validated by a listening study conducted in \cite{nayem2021towards}.  With parameter $r=100$, the total number of quantization levels, $\mathcal{D}$, is $1600$. The QSM tunable parameter, $\mu$, is set to $0.01$.


\section{Results}
\label{sec:results}

\subsection{MOS prediction results}
\label{subsec:mos_results}
We first evaluate our MOS-prediction performance in comparison with other approaches. In particular, we compare against NISQA~\cite{mittag2019non}, which we modified to estimate human-accessed MOS. Originally, they estimate perceptual objective listening quality assessment (POLQA)~\cite{beerends2013perceptual} scores using a CNN and BLSTM architecture. We also compare against the PMOS model proposed in~\cite{dong2020pyramid}, which is identical in structure to our PMOS model. Finally, we include our proposed SE+PMOS approach~\cite{nayem2021incorporating} (no joint training), where our PMOS model is held fixed while the SE model is training using the embeddings from the PMOS encoder. 

We use four metrics to evaluate MOS-estimation performance: mean absolute error (MAE), epsilon insensitive root mean squared error (RMSE)~\cite{rec2012p}, Pearson’s correlation coefficient $\gamma$ (PCC), and Spearman’s rank correlation coefficient $\rho$ (SRCC). 


\begin{table}[t!]

\centering
\caption{Performance comparison with MOS prediction models {comparing against the ground truth MOS obtained from human subjects}. Best results are shown in \textbf{bold}.}
\label{tab:mos_results}
\resizebox{\columnwidth}{!}{%
\begin{tabular}{| l | c c c c | }
\cline{2-5}
   \multicolumn{1}{c|}{}         & {MAE}$\downarrow$ & {RMSE}$\downarrow$ & {PCC ($\gamma$)}$\downarrow$ & {SRCC ($\rho$)}$\downarrow$ \\ \hline
   
NISQA~\cite{mittag2019non}    & 0.62 ($\pm$0.18)        & 0.7 ($\pm$0.16)      & 0.71 ($\pm$0.14)           & 0.79 ($\pm$0.15)            \\
PMOS~\cite{dong2020pyramid}                      & 0.51 ($\pm$0.15)         & 0.57 ($\pm$0.12)          & 0.88 ($\pm$0.17)           & 0.88 ($\pm$0.14)           \\
SE+PMOS~\cite{nayem2021incorporating}                     & \textbf{0.45} ($\pm$0.08) & \textbf{0.52} ($\pm$0.09) & \textbf{0.9} ($\pm$0.12) & \textbf{0.91} ($\pm$0.1)           \\
Proposed                     & \textbf{0.45} ($\pm$0.08) & \textbf{0.52} ($\pm$0.09) & \textbf{0.9} ($\pm$0.12) & \textbf{0.91} ($\pm$0.1)         \\
\hline
\end{tabular}
}
\end{table}

Table~\ref{tab:mos_results} shows the results, where our proposed approach and SE+PMOS clearly outperform the other MOS prediction models according to all metrics. MAE is minimized by $0.6$ compared to the original PMOS~\cite{dong2020pyramid} approach. There is also a $0.05$ reduction in RMSE. This justifies our proposed approach that combines MOS estimation and speech enhancement tasks. Note, however, that similar results are obtained for our proposed approach and the SE+PMOS approach, which suggests that joint training (e.g., fine tuning) may help speech enhancement more than MOS prediction.

\subsection{Speech enhancement model}
\label{subsec:se_results}

\begin{table*}[t!]
\centering
\caption{Average results of the speech enhancement models in different performance metrics. Best results are shown in \textbf{bold}.}
\label{tab:results_cosineVoices}
\resizebox{\linewidth}{!}{%
\begin{tabular}{ | l | l | c c c c | c c c c | }
\cline{3-10}
\multicolumn{1}{l}{\multirow{2}{*}{}} &                    & \multicolumn{4}{ c |}{{COSINE}}                             & \multicolumn{4}{ c |}{{VOiCES}} 
\\ \hline
\multicolumn{1}{|l|}{{models}}                 & \multicolumn{1}{c|}{{loss func.}} & {PESQ}$\uparrow$ & {SI-SDR}$\uparrow$ & {ESTOI}$\uparrow$ & {MOS-LQO}$\uparrow$ & {PESQ}$\uparrow$ & {SI-SDR}$\uparrow$ & {ESTOI}$\uparrow$ & {MOS-LQO}$\uparrow$ \\ \hline
\multicolumn{1}{|l|}{{Mixture}}                                        & {-}                                                       & {1.46} & {0.53}   & {0.62}  & {4.04}    & {1.26} & {-1.3}   & {0.48}  & {2.74}    \\ \hline
\multicolumn{1}{|l|}{}                                                        & mse                                                              & 2.68          & 2.8             & 0.8            & 3.2              & 2.3           & 1.2             & 0.69           & 3.5              \\ 
\multicolumn{1}{|l|}{}                                                        & mos~\cite{fu2019learning}                                                              & 2.8           & 3.8             & 0.82           & 4.2              & 2.37          & 1.66            & 0.74           & 5.3              \\ 
\multicolumn{1}{|l|}{}                                                        & mse+sa                                                           & 2.72          & 3.1             & 0.82           & 4                & 2.35          & 1.6             & 0.7            & 3.8              \\ 
\multicolumn{1}{|l|}{}                                                        & mos+sa                                                           & 2.89          & 4.1             & 0.85           & 4.4              & 2.42          & 1.72            & 0.77           & 5.7              \\ 
\multicolumn{1}{|l|}{\multirow{-5}{*}{SE}}                                    & sdr~\cite{kawanaka2020stable}                                                              & 2.7           & 4.5             & 0.82           & 3.4                & 2.32          & 2.01            & 0.72           & 3              \\ \hline
\multicolumn{1}{|l|}{{ }}                                 & mse                                                              & 3.1           & 4               & 0.85           & 4.2              & 2.48          & 1.8             & 0.8            & 6                \\ 
\multicolumn{1}{|l|}{{}}                                 & mse+sa                                                           & 3.19          & 4.6             & 0.93           & 4.8              & 2.54          & 2.08            & 0.86           & 6.3              \\  
\multicolumn{1}{|l|}{\multirow{-3}{*}{SE+PMOS~\cite{nayem2021incorporating}}}        & mse+sa+mos                                                       & 3.19          & 4.5             & 0.92           & \textbf{5.1}     & 2.53          & 2.06            & 0.84           & \textbf{6.5}     \\ \hline
\multicolumn{1}{|l|}{}                                                        & pesq                                                             & \textbf{3.28} & 4.4             & 0.9            & 5                & \textbf{2.67} & 2.01            & 0.83           & 6.1              \\ 
\multicolumn{1}{|l|}{\multirow{-2}{*}{MetricGAN~\cite{fu2019metricGAN}} }                             & stoi                                                             & 3.19          & 4.3             & \textbf{0.94}  & 4.8              & 2.5           & 2               & \textbf{0.87}  & 5.8              \\ \hline
\multicolumn{1}{|l|}{SSEMS~\cite{zezario2019specialized}}                                                   & qnet ($\phi=0dB$)                                                       & 2.85          & 2.99            & 0.83           & 3                & 2.4           & 1.8             & 0.7            & 2.8              \\ \hline
\multicolumn{1}{|l|}{{Chi++\textsubscript{fQSM,bS}~\cite{nayem2021towards}}}                     &    dc+cls+sa                                                              & 2.9           & 3.3             & 0.84           & 3.4              & 2.44          & 1.78            & 0.7            & 3                \\ \hline
\multicolumn{1}{|l|}{}                                 & mse+sa                                                           & 3.25          & 4.8             & \textbf{0.94}  & 4.75             & 2.64          & 2.1             & \textbf{0.87}  & 6.2              \\ 
\multicolumn{1}{|l|}{\multirow{-2}{*}{Proposed}} & mse+sa+mos                                                       & 3.25          & \textbf{4.82}   & \textbf{0.94}  & 5.04             & 2.64          & \textbf{2.13}   & \textbf{0.87}  & 6.47             \\ \hline
\end{tabular}
}
\end{table*}
For speech enhancement, we compare against a baseline approach without an attention mechanism \cite{graves2013speech}. We denote this baseline approach as SE. Five separate loss functions are applied to optimize this approach, and they are MSE, MSE plus signal approximation, MOS, signal approximation with MOS, and SDR. To compute the MOS loss function, we utilize the SE loss function from \cite{fu2019learning} which leverages objective-MOS (oMOS) ratings learned from a speech assessment model~\cite{fu2018quality}. SDR~\cite{kawanaka2020stable} loss functions are proposed in literature previously with different enhancement architectures. For the SDR loss function, the SE model is optimized using the following cost function:
\begin{align}
    \mathcal{L}_{SDR} = \sum_{n=1}^N \mathcal{K}_{\theta}  \Big( 10 \log \frac{\Vert s^n\Vert^2}{\Vert s^n-\hat{s}^n\Vert^2} \Big)
\end{align}
where $\mathcal{K}_\theta(a)=\theta\cdot \tanh(\frac{a}{\theta})$, $\theta$ is a clipping parameter, $N$ is the mini-batch size, and $s^n$ and $\hat{s}^n$ are the n\textsuperscript{th} sample of the clean and estimated speech signal in time. We use $\theta=20$ in our training. We also compare against a generative adversarial network (GAN) approach that individually optimizes with PESQ and STOI~\cite{fu2019metricGAN}. We denote this model as MetricGAN. 
They estimate the IRM for a speech mixture conditioned on a GAN discriminator that outputs evaluation scores in continuous space (i.e. scores between 0 and 1) based on either normalized PESQ or STOI target metrics. 
We compare our model with the ensemble-based Specialized Speech Enhancement Model Selection (SSEMS) approach~\cite{zezario2019specialized} that uses Quality-Net~\cite{fu2018quality} as its objective function in a black-box manner. Quality-Net is an oMOS approach that estimates the Perceptual Evaluation of Speech Quality (PESQ) score. The SSEMS approach uses an ensemble of enhancement models, each trained on audio at specific SNRs and speaker genders. During inference, it selects the output with the highest PESQ score. SSEMS uses a SNR threshold of $20$ dB, while we use a threshold of $0$ dB for balanced training and better performance. Additionally, we conduct a comparison with our initial approach that integrates MOS embeddings in speech enhancement, as presented in \cite{nayem2021incorporating}. This model is referred to as SE+PMOS, and it does not involve joint training or the QSM language model. We evaluate SE+PMOS with varying combinations of loss functions. 
All models are trained using the experimental setup that is previously mentioned. We modify the comparison models using the code provided by the original authors.

We assess speech enhancement performance using PESQ~\cite{rix2001perceptual}, scale-invariant SDR (SI-SDR)~\cite{le2019sdr}, and extended STOI (ESTOI)~\cite{jensen2016algorithm}. In the absence of actual human quality objective, we measure the predicted MOS score of the enhanced speech, using our proposed PMOS model, since we aim to improve human-assessed speech quality. We denote this metric as MOS listener quality objective (MOS-LQO). Table~\ref{tab:results_cosineVoices} shows the average results of the different enhancement models, according to each of the performance metrics on COSINE and VOiCES dataset. As the scores of the unprocessed mixtures show, the VOiCES corpus is  more challenging than the COSINE corpus. 
With the baseline SE model, we experiment with 5 different combination of loss functions. Using the MSE loss only in SE:mse, we see improvements in objective scores, except with MOS-LQO for the COSINE data. Then we apply a MOS loss $\mathcal{L}_{mos}$ as the sole objective criterion, as proposed in \cite{fu2019learning}. Our experimental results show that this approach results in an overall improvement of $1.4$ in MOS-LQO compared to SE:mse. 
Then we separately combine the signal approximation loss with the mse loss and MOS loss (e.g., mse+sa and mos+sa). In PESQ, we gain an average of $\ge0.05$ and $\ge0.07$ compared to the models that use only the MSE loss and only the MOS loss, respectively. Furthermore, the model trained with the mos+sa loss function achieves the highest MOS-LQO score of $4.4$ and $5.7$ among all five loss functions tested with the SE model in COSINE and VOiCES dataset, respectively. This result is on average $1.15$ MOS-LQO higher than that obtained with the mse+sa loss function. These scores suggest that $\mathcal{L}_{mse}$ and $\mathcal{L}_{sa}$ maximize the overall speech intelligibility, whereas $\mathcal{L}_{mos}$ guides the model towards perceptual speech quality. Note that in all these $\mathcal{L}_{mos}$ calculations, we use a separately trained PMOS model's output without joint learning.
Lastly, we apply the SDR loss function as proposed in \cite{kawanaka2020stable}, which is used as the pre-training stage for model training. We observe an average gain of $0.9$ in SI-SDR, however, it yields a poor score according to other metrics, especially a $0.7$ loss in MOS-LQO compared to SE with mse and sa loss terms. 

SE+PMOS is separately investigated with 3 combinations of loss functions, i.e. mse, mse+sa, and mse+sa+mos. Compared with SE models, SE+PMOS with mse loss achieves $0.9$ SI-SDR and $1.75$ MOS-LQO improvements on average, which shows the benefit of incorporating the PMOS model. The SE+PMOS:mse+sa model improves the performance further with an average of $0.14$ ESTOI gain over the SE:mse+sa model. The inclusion of the mos loss gives the best MOS-LQO scores of $5.1$ and $6.5$ over all the comparison models in noisy and reverberant conditions, respectively.


\begin{table*}[t!]
\centering
\caption{Average testing results of the speech enhancement models on CHiME-5 and CHiME-4 datasets. Best results are shown in \textbf{bold}.}
\label{tab:results_chime}
\resizebox{\linewidth}{!}{%
\begin{tabular}{| l | l | c c c c c | c c c c c |}
\cline{3-12}
\multicolumn{1}{l}{\multirow{2}{*}{}} &                    & \multicolumn{5}{ c |}{{CHiME-5}}                             & \multicolumn{5}{ c |}{{CHiME-4}} 
\\ \hline
\multicolumn{1}{|l|}{models}                          & \multicolumn{1}{c|}{loss func.} & PESQ$\uparrow$          & SI-SDR$\uparrow$       & ESTOI$\uparrow$         & MOS-LQO$\uparrow$      & WER\%$\downarrow$         & PESQ$\uparrow$          & SI-SDR$\uparrow$        & ESTOI$\uparrow$         & MOS-LQO$\uparrow$      & WER\%$\downarrow$         \\ \hline
\multicolumn{1}{|l|}{Mixture}                & -                               & 1.7           & 2.4          & 0.52          & 3.8          & 152.1         & 1.96          & 2.86          & 0.6           & {4.6} & {33.7} \\ \hline
\multicolumn{1}{|l|}{SE}                              & mos+sa                          & {2.25} & {3.9} & {0.62} & {4}   & {96.4} & {2.32} & {5.22} & {0.63} & {5}   & {25.6} \\ \hline
\multicolumn{1}{|l|}{SE+PMOS}                         & mse+sa+mos                      & 2.37          & 6.1          & 0.67          & 4.4          & 84.5          & 2.45          & 7.6           & 0.7           & 5.8          & 22.6          \\ \hline
\multicolumn{1}{|l|}{\multirow{2}{*}{MetricGAN}}      & pesq                            & \textbf{2.44} & {6.3} & {0.65} & {4.1} & {94.8} & \textbf{2.51} & {7}    & {0.68} & {5.3} & {19.7} \\ 
\multicolumn{1}{|l|}{}                                & stoi                            & 2.39          & 6.2          & \textbf{0.71} & 4.1          & 91.3          & 2.45          & {6.45} & \textbf{0.73} & 5.6          & 21.5          \\ \hline
\multicolumn{1}{|l|}{\multirow{2}{*}{Proposed}} & mse+sa                          & 2.41          & 7.1          & {0.68} & 4.7          & \textbf{78.3} & 2.5           & {7.9}  & 0.72          & 5.76         & \textbf{18.1} \\ 
\multicolumn{1}{|l|}{}                                & mse+sa+mos                      & 2.41          & \textbf{7.3} & {0.68} & \textbf{4.9} & 79.4          & {2.5}  & \textbf{8.61} & \textbf{0.73} & \textbf{6}   & 18.9          \\ \hline
\end{tabular}}
\end{table*}
MetricGAN optimizes PESQ or STOI, therefore, it outperforms other comparison models in terms of PESQ and ESTOI, although the scores for the SE+PMOS approaches are higher according to the other evaluation metrics even though these metrics are not leveraged during training. 
SSEMS yields the lowest scores across all metrics compared with SE+PMOS and MetricGAN approaches, though we do parameter tuning for this model.
Chi++\textsubscript{fQSM,bS} estimates quantized speech, and the results show that it affects the traditional objective functions. This performs poorly compared with the SE+PMOS and MetricGAN approaches, however, on average, it outperforms SSEMS in all criteria, and the SE models in terms of PESQ. With the MOS-LQO criteria, it fails to produce good scores. This points out the importance of incorporating perceptual features during enhancement, which Chi++\textsubscript{fQSM,bS} clearly lacks.

We calculate the performance of our proposed model using two combinations of loss functions. 
Using only mse and sa loss terms, we achieve the highest ESTOI scores for both corpora, though these results are nearly identical to the model trained with all three loss terms. Using $\mathcal{L}$ (eq:\ref{eq:loss}) in our proposed model, we obtain the highest SI-SDR scores while maintaining similar PESQ and ESTOI performance as compared to the best-performing model. Specifically, our proposed model achieves the highest ESTOI score and an average PESQ score that is only $0.03$ less than that of the best performing MetricGAN:pesq model.
Contrasting with the Chi++\textsubscript{fQSM,bS} model, which uses spectral language model to estimate quantized speech, our proposed approach outperforms the quantized model according to all metrics, which proves the significance of joint learning.
When comparing MOS-LQO scores, our proposed:mse+sa+mos model achieves better scores than the other models except the SE+PMOS:mse+sa+mos model with an average of only $0.05$ declination. Thus, the inclusion of a spectral language model helps the model proposed (e.g., mse+sa+mos) to estimate better quality speech according to the overall evaluation criteria. 
It is important to note that our proposed approach performs best according to SI-SDR in both noisy and reverberant environments, where this metric is not used by any of the approaches during optimization.  

We further examine our approaches using completely unseen corpora. We test models with the CHiME-5 and CHiME-4 corpora where the models are trained from the COSINE dataset according to the system setup mentioned in section~\ref{subsec:setup}. Table~\ref{tab:results_chime} shows the performance evaluated according to PESQ, SI-SDR, ESTOI, MOS-LQO, and word error rate (WER). To calculate WER, we use the conventional ASR baseline that is provided with CHiME-5 and CHiME-4 dataset. We investigate WER with both GMM based ASR and end-to-end ASR, however, we find that the end-to-end approach results in a higher error compared to the GMM baseline. This might happen due to larger data requirements of the end-to-end ASR system as mentioned in \cite{barker2018fifth}. Therefore, we use the GMM ASR approach to compare the WER performance of the enhancement models.
From the scores of mixtures, we find that CHiME-5 is more challenging than CHiME-4 with a $118.8\%$ higher WER and a $0.46$ lower SI-SDR. Our proposed approach yields the best MOS-LQO scores with $4.9$ with CHiME-5 and $6$ with CHiME-4 data. The proposed mse+sa model results in the lowest WER of $78.3$ and $18.1$ using CHiME-5 and CHiME-4, respectively. Note that the WER of the GMM baseline ASR for the CHiME-5 challenge is $72.8$ in binaural and $91.7$ in single array conditions. Here our approaches enhance monaural speech, a more challenging condition. Our proposed approach outperforms other comparison models in terms of SI-SDR with a $5.29$ average improvement compared to others. According to PESQ and ESTOI metrics, MetricGAN variants give the best performace, however, proposed model's performance is $0.02$  and $ 0.015$ lower according to PESQ and ESTOI, respectively, for the best performing MetricGAN models. Hence, our proposed approach is effective on out-of-vocabulary scenario trained by a comparable dataset.



   


\begin{figure}[b!]
    \centering
\begin{tikzpicture}
	\begin{axis}[
	    cycle list/Dark2-4,
		boxplot/draw direction = y,
		boxplot/box extend=0.8,
		axis x line* = bottom,
		axis y line = left,
		enlarge y limits,
		ymajorgrids,
		xtick = {1, 2, 3, 4, 5, 6, 7, 8},
		xticklabel style = {align=center, font=\small, rotate=60, alias={xtick-\ticknum}},
		xticklabels = {Mixture, SE+PMOS, MetricGAN, Proposed, Mixture, SE+PMOS, MetricGAN, Proposed},
		ylabel = {MOS},
		ytick = {1, 2, 3, 4, 5},
	]
	
	\addplot+[
        boxplot prepared={
        lower whisker=1, lower quartile=1.45,
        median=1.74,
        upper quartile=2.5, upper whisker=4.05, }, fill, draw=black]
        coordinates {}
        node[above, color=black] at
        (boxplot box cs: \boxplotvalue{median},.5)
        {\scriptsize \pgfmathprintnumber{\boxplotvalue{median}}};
    \addplot+[
        boxplot prepared={
        lower whisker=1.38, lower quartile=1.84,
        median=2.28,
        upper quartile=3.1, upper whisker=4.3, }, fill, draw=black]
        coordinates {}
        node[above, color=black] at
        (boxplot box cs: \boxplotvalue{median},.5)
        {\scriptsize \pgfmathprintnumber{\boxplotvalue{median}}};
    \addplot+[
        boxplot prepared={
        lower whisker=1.3, lower quartile=1.75,
        median=2.13,
        upper quartile=3.2, upper whisker=4.1, }, fill, draw=black]
        coordinates {}
        node[above, color=black] at
        (boxplot box cs: \boxplotvalue{median},.5)
        {\scriptsize \pgfmathprintnumber{\boxplotvalue{median}}};
    \addplot+[
        boxplot prepared={
        lower whisker=1.4, lower quartile=1.9,
        median=2.46,
        upper quartile=3.16, upper whisker=4.34, }, fill, draw=black]
        coordinates {}
        node[above, color=black] at
        (boxplot box cs: \boxplotvalue{median},.5)
        {\scriptsize \pgfmathprintnumber{\boxplotvalue{median}}};
        
    \addplot+[
        boxplot prepared={
        lower whisker=1.0, lower quartile=1.35,
        median=1.64,
        upper quartile=2.39, upper whisker=4.18, }, fill, draw=black]
        coordinates {}
        node[above, color=black] at
        (boxplot box cs: \boxplotvalue{median},.5)
        {\scriptsize \pgfmathprintnumber{\boxplotvalue{median}}};
    \addplot+[
        boxplot prepared={
        lower whisker=1.31, lower quartile=1.8,
        median=2.18,
        upper quartile=2.76, upper whisker=4.24, }, fill, draw=black]
        coordinates {}
        node[above, color=black] at
        (boxplot box cs: \boxplotvalue{median},.5)
        {\scriptsize \pgfmathprintnumber{\boxplotvalue{median}}};
    \addplot+[
        boxplot prepared={
        lower whisker=1.26, lower quartile=1.71,
        median=2.06,
        upper quartile=3.17, upper whisker=4.32, }, fill, draw=black]
        coordinates {}
        node[above, color=black] at
        (boxplot box cs: \boxplotvalue{median},.5)
        {\scriptsize \pgfmathprintnumber{\boxplotvalue{median}}};
    \addplot+[
        boxplot prepared={
        lower whisker=1.34, lower quartile=1.85,
        median=2.25,
        upper quartile=3.07, upper whisker=4.48, }, fill, draw=black]
        coordinates {}
        node[above, color=black] at
        (boxplot box cs: \boxplotvalue{median},.5)
        {\scriptsize \pgfmathprintnumber{\boxplotvalue{median}}};
        
	\end{axis}
	
	\path (0,0) coordinate (P);
    \draw [thick,decoration={brace,mirror,raise=5em},decorate] (xtick-0|-P) -- (xtick-3.5|-P) 
        node[midway,yshift=-6em]{CHiME-4};
    \draw [thick,decoration={brace,mirror,raise=5em},decorate] (xtick-4|-P) -- (xtick-7.5|-P) 
        node[midway,yshift=-6em]{CHiME-5};


\end{tikzpicture}

\caption{MOS ratings of the speech enhancement modes on CHiME-4 and CHiME-5 datasets using DNSMOS P.835.}
\label{fig:dnsmos_results}
\end{figure}
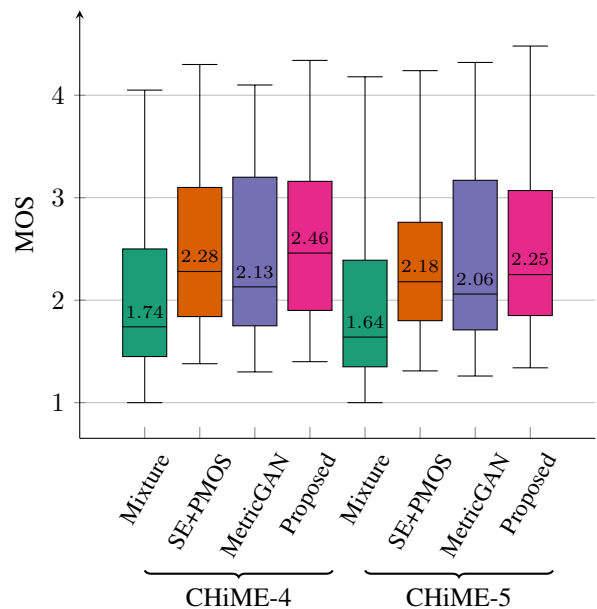

\subsection{Perceptual quality evaluation}
\label{subsec:dnsmos}

We finally evaluate our model using P.835 metric~\cite{reddy2022dnsmos} to measure perceptual quality. We calculate the DNSMOS score on a scale of $[1-5]$ ($1$ = worst, $5$ = best) for the mixture, PMOS+SE, MetricGAN, and our proposed models using the CHiME-4~\cite{vincent2017analysis} and CHiME-5~\cite{barker2018fifth} datasets (simulated and real-recording). Figure~\ref{fig:dnsmos_results} shows the scores. With CHiME-4, the original mixture scores range from $1.45$ to $2.5$ with a median of $1.74$. Our proposed model achieves a median MOS of $2.46$, which is higher than the others. Fon CHiME-5, the original mixture scores range from $1.0$ to $4.18$. Our proposed model outperforms the others with a median of $2.25$. Our proposed model and PMOS+SE have smaller standard deviations compared to MetricGAN. Overall, our proposed model improves noisy speech in both the acoustic and perceptual aspects.

\section{Discussion}
\label{sec:discuss}

Our proposed model outperforms all comparison models on SI-SDR metrics for both seen and unseen datasets, without optimization of any of the models (Table \ref{tab:results_cosineVoices}, \ref{tab:results_chime}). This means that our approach improves speech quality by minimizing the distortion ratio when separated from the noise component. Additionally, our models yield the best MOS-LQO ratings on real-world captured audios (CHiME datasets, Table \ref{tab:results_chime}). These results are consistent with the findings of \cite{zezario2022deep, nayem2021incorporating} that incorporating embeddings from a speech assessment model improves SE performance, and the results of \cite{braun2022effect} that using MOS loss during model optimization leads to higher MOS-LQO scores. Our proposed approach achieves PESQ and ESTOI scores that are only slightly lower than those of the best-performing model, with a difference of only $0.03$ and $0.01$, respectively. This indicates that speech quality and intelligibility metrics are closely related to the subjective speech quality metric (MOS-LQO), and that these metrics can be improved without explicit optimization. Furthermore, our proposed model achieves the best average DNSMOS scores with low standard deviations on CHiME datasets (Figure \ref{fig:dnsmos_results}), indicating that it is effective in a wide range of real-world noise levels. This is a desirable quality for an effective SE model to be effective not only in high SNRs and limited noisy environments, but also in large SNR ranges and real-world conditions such as those offered by the CHiME dataset.

When comparing our proposed model that uses mse+sa+mos loss to the PMOS+SE model (as shown in Table \ref{tab:results_chime}), we can observe significant improvements in all performance metrics. As both models use the same loss function, the improvements are attributed to the incorporation of LM and the joint learning method. Moreover, we found that these two models exhibit similar performance on the MOS prediction (Table \ref{tab:mos_results}), indicating that the benefits of joint learning mostly impact the enhancement part of the model.

An intriguing finding is that our proposed model shows a decline in WER\% when MOS loss is incorporated, especially for larger real-world recordings such as CHiME-5, with degradation up to $1.1$. Although our study is not primarily concerned with ASR performance, this suggests a potential trade-off between ASR accuracy and subjective speech quality scores. Further investigation is needed to comprehend this relationship.

Our proposed method demonstrates that training a speech enhancement (SE) model and a MOS-based speech assessment model jointly can lead to better speech quality measured by objective metrics such as perceptual quality, intelligibility, and MOS ratings. However, we acknowledge that our study's use of subjective MOS (sMOS) estimation instead of actual human listeners may introduce discrepancies between MOS-LQO and human-rated MOS, which could impact our findings. To address this limitation, we plan to conduct sMOS evaluation by human listeners in future work. Although we used the same MOS prediction model for all comparison models, we believe that incorporating human-rated sMOS evaluations will provide more robust insights into our proposed method's effectiveness.
For computing loss terms, we opt for the MSE loss function along with a bi-gram language model that considers only time-along transitions. Our aim is to keep the model simple and focus on the effectiveness of our approach. However, we acknowledge that using different loss functions for different loss components and employing a more complex language model that considers both temporal and spectral transition levels can be beneficial. We plan to explore these possibilities in our future work.


\section{Conclusion}
\label{sec:conclusion}
Our proposed speech enhancement model utilizes a speech quality MOS assessment metric in a joint learning manner and incorporate quantized ASR-style language model for better performance. The results show that it outperforms other models in both noisy and reverberant environments, as well as in unseen real-world noisy conditions. It shows that perceptually-relevant embeddings are useful for speech enhancement. However, we evaluate our model's subjective score using a MOS-estimation model. Additionally, our assessment model provides utterance-level feedback, which may be sub-optimal since the model's embeddings are calculated at the frame level. In our proposed LM, we consider only bi-gram spectral models which are generated by considering only along-time transitions.
In the future, we will explore higher-order N-gram models that consider both temporal and spectral transitions to enhance both magnitude and phase responses. We will address per-frame or window level perceptual score generation in future work.

%
\bibliographystyle{IEEEtran}
\bibliography{0_main.bbl}

%
\begin{IEEEbiography}
[{\includegraphics[width=1in,height=1.25in,clip,keepaspectratio]{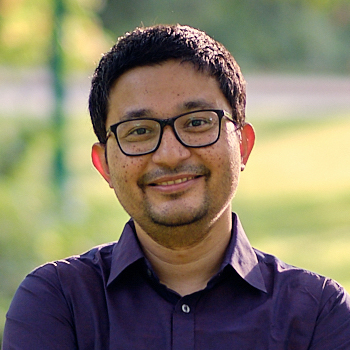}}]{Khandokar Md. Nayem} (Student Member, IEEE)
received the B.Sc. degree in computer science and engineering from the Bangladesh University of Science and Engineering, Dhaka, Bangladesh, in 2014 and the M.Sc. degree in computer science from the Indiana University, Bloomington, IN, USA, in 2019, where he is currently working toward the Ph.D. degree in computer science. His research interests include speech enhancement/processing, deep learning, and human speech perception.
\end{IEEEbiography}

\begin{IEEEbiography}
[{\includegraphics[width=1in,height=1.25in,clip,keepaspectratio]{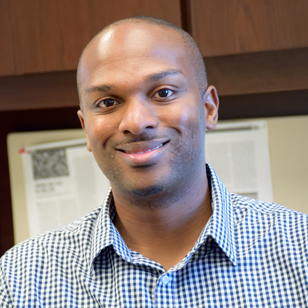}}]{Donald S. Williamson} (Senior Member, IEEE)
received the B.E.E. degree in electrical engineering from the University of Delaware, Newark, DE, USA, the M.S. degree in electrical engineering from Drexel University, Philadelphia, PA, USA, and the Ph.D. degree in computer science and engineering from The Ohio State University, Columbus, OH, USA. He is currently an Associate Professor with the Department of Computer Science and Engineering, Ohio State University, Columbus, OH, USA. His research interests include speech enhancement/separation, speech assessment, and audio privacy.
\end{IEEEbiography}

\end{document}